\documentclass[final,12pt,3p]{elsarticle}
\usepackage{latexsym}
\usepackage{amssymb}
\usepackage{amsmath}
\usepackage{cancel}
\usepackage{color}
\usepackage{tikz}
\definecolor{red}{rgb}{1.0,0.0,0.0}
\definecolor{blue}{rgb}{0.0,0.0,1}
\definecolor{green}{rgb}{0.29, 0.33, 0.13}
\newcommand{\ket}[1]{\left| #1 \right>} 
\newcommand{\bra}[1]{\left< #1 \right|} 
\newcommand{\meanv}[1]{\left< #1 \vphantom{#1} \right>} 
 
\newcommand{\abs}[1]{\left| #1 \vphantom{#1} \right|} 
\newcommand{\pms}{\mathbin{\tikz[x=1.4ex,y=1.4ex,line width=.1ex] \draw (0.0,0) -- (1.0,0) (0.5,0.08) -- (0.5,0.92) (0.0,0.5) -- (1.0,0.5);}}%
\newcommand{\p}{\mathbin{\tikz[x=1.4ex,y=1.4ex,line width=.1ex] \draw (0.0,0) -- (0.0,0) (0.5,0.08) -- (0.5,0.92) (0.0,0.5) -- (1.0,0.5);}}%
\newcommand{\m}{\mathbin{\tikz[x=1.4ex,y=1.4ex,line width=.1ex] \draw (0.0,0) -- (1.0,0) ;}}
\begin{document}
\begin{frontmatter}
\title{The strange attraction phenomenon in cQED: the intermediate quantum coupling regime}
%
\author[label1]{Santiago Echeverri-Arteaga}
\author[label1]{Herbert Vinck-Posada}
\address[label1]{Departamento de F\'isica, Universidad Nacional de Colombia, 111321, Bogot\'a, Colombia}
\author[label2]{Edgar A. G\'omez\corref{cor1}}
\address[label2]{Programa de F\'isica, Universidad del Quind\'io, 630004, Armenia, Colombia\fnref{label4}}
\cortext[cor1]{Corresponding author}
\ead{eagomez@uniquindio.edu.co}
\begin{abstract}
We theoretically investigate the strange phenomenon of attraction between the cavity mode and the exciton mode that has been experimentally observed in a photonic crystal cavity. Our results within the Lindblad master equation approach successfully explain why the apparent cavity-to-exciton spectral shifting occurs, and how this phenomenon is a manifestation of a dynamical phase transition in the system. Moreover, our findings are in good qualitative agreement with pioneer experimental results as well as support the hypothesis that an intermediate coupling regime exists in the cavity quantum electrodynamics. In fact, this new quantum regime exhibits phenomenology from both weak and strong coupling regimes simultaneously.
\end{abstract}
\end{frontmatter}
%
\section{Introduction}\label{intro}
\noindent
Within the framework of the cavity quantum electrodynamics (cQED), an extensive experimental and theoretical efforts have been achieved resulting in the unequivocal proof of the existence of the weak and strong coupling regime \cite{Reithmaier:2004,Shudong:1998}. Thus, significant progress has been made in the characterization of the role played by the different quantum regimes as well as of its applications. For example, the weak coupling regime has been harnessed for fabricating organic light-emitting devices (OLEDs) \cite{Bulovic:1998,Fletcher:2000} and  the generation of entangled photons 
for optical quantum computing \cite{Salter:2010}. In the case of the strong coupling regime, it has been useful for designing and realization of single photon emitters \cite{Moreau:2001,Tsintzos:2008} and promising devices based on solid-state quantum systems with a high scalability for quantum computation \cite{Wei:2014}. With the current state of technology, it is possible to achieve a control of these quantum regimes where applications in quantum plasmonics and metamaterials are expected \cite{Ginzburg:2016}, as well as in photonic nanostructures for developing physical platforms for the quantum information science \cite{Kiraz:2003}. All this promising technology is based on our understanding of the fundamental physics laws that govern the light-matter interaction in a domain close enough to the resonance. Despite the success achieved by the cQED, and more precisely for describing the underlying physics of quantum dots (QDs) coupled to semiconductor cavities, many fundamental questions remain open when the off-resonant QD-cavity coupling is considered. In fact, an amazing as well as inexplicable spectral shifting (cavity-to-exciton attraction) of the cavity mode toward to the exciton mode has been observed experimentally by Tawara {\it et al.}~\cite{Tawara:2010}. This pioneer work has suggested the existence of a new intermediate coupling regime which remains almost unexplored up to date. This interesting quantum regime could give rise to new technological developments. More recently, experiments involving a single QD as well as biexcitons embedded in a micropillar cavity have confirmed that this phenomenon is real and not a mere experimental artifact~\cite{Valente:2014}. Surprisingly and despite of the active research in cQED systems, this quantum phenomenon has not yet attracted much attention as was done by another off-resonance phenomenon in cQED \cite{Hennessy:2007}. Consequently, there are a very few theoretical works devoted to explain the origin of this intriguing phenomenon which has been attributed to the effects due to the pure dephasing or phononic mechanisms that mainly participate to the cavity feeding~\cite{Kaniber:2008}. In this paper, we explain clearly the attraction phenomenon that has been reported experimentally, and moreover how this new quantum regime can be identified with a dynamical phase transition in the system \cite{Echeverri-Arteaga:2017}. This paper is organized as follows. In Section~\ref{sec:background}, we present the theoretical model for describing a quantum dot-cavity system together with a description of the cavity-to-exciton attraction phenomenon through the full Lindblad master equation approach. Additionally, in this section we present a simplified model based on the Lindblad master equation approach without gain for understanding the relationship between the dynamical phase transition and this unexpected quantum phenomenon. Finally, the conclusions are summarized in Section~\ref{sec:conclusions}.
\section{Theoretical background and discussion}\label{sec:background}
\subsection{Description based on full Lindblad master equation approach.}\label{Fulllindblad}
\begin{figure}[h!]
\centering
\includegraphics[scale=1]{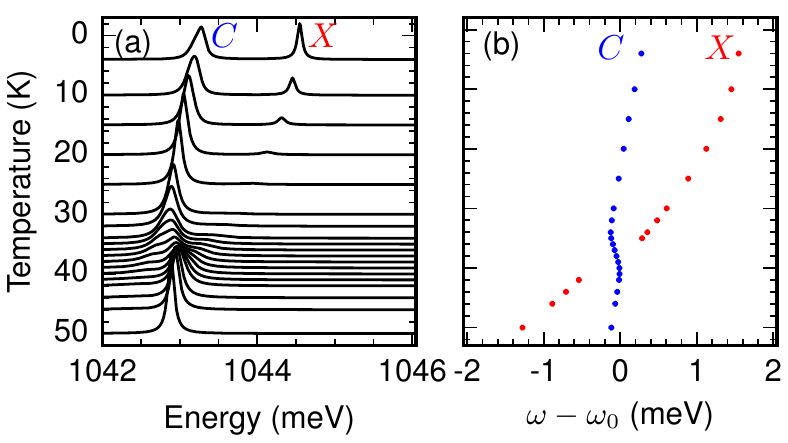}
\caption{(a) PL spectra as a function of the temperature. The emission peaks labeled as $C$ and $X$ approaches one another near the crossover energy $\omega_0\approx1042.94\,meV$ at  temperature $T\approx37.43\,K$. (b) The spectral peak positions clearly shows that the cavity mode blueshift (blue dots) toward the exciton mode (red dots). The parameters chosen in our numerical simulations are: $g=0.3\,meV$, $\kappa=0.1\,meV$, $\gamma_x=0.001\,meV$, $P_x=0.06\,meV$. We have fixed for the phonon decay rate $P_{\theta}$ the values: $A=0.5$, $B=0.2\,K^{-1}$, $T'=30\,K$ and $\tilde{P}_{\theta}=0.45\,meV$. (For interpretation of the references to color in this figure legend, the reader is referred to the web version of this article.)
}\label{Fig-Tawara}
\end{figure}

In order to elucidate the strange attraction phenomenon observed experimentally in a planar photonic-crystal cavity (PhCC) with an embedded single QD, we take advantage of the well-known Jaynes-Cummings (JC) Hamiltonian, which reads ($\hbar=1$), $\hat{H}=\omega_{x}\hat{\sigma}^{\dag}\hat{\sigma}+\omega_c\hat{a}^{\dag}\hat{a}+g(\hat{a}^\dag\hat{\sigma}+\hat{a}\hat{\sigma}^\dag)$ with $g$ the light-matter interaction constant between the cavity mode and the QD. Additionally, $\hat{a}$ and $\hat{\sigma}=\ket{0}\bra{1}$ are the annihilation and lowering operators for the cavity mode and the QD, respectively. Notice that the lowering operator acts from the excited state $\ket{1}$ to the ground state $\ket{0}$, moreover the detuning between the QD and the cavity mode is defined as $\Delta=\omega_{x}-\omega_c$. Here $\omega_{x}$ and $\omega_{c}$ are the frequencies associated to the QD and the cavity mode, respectively. Besides, we incorporate the influence of the environment by considering the irreversible processes of leakage of photons from the cavity at rate $\kappa$, the spontaneous emission $\gamma_x$, the incoherent QD pumping $P_x$ and a phonon-mediated QD-cavity coupling $P_\theta$ through the Lindblad master equation
\begin{eqnarray}\label{eq1}
\frac{d\hat{\rho}}{dt}&=&-i[\hat{H},\hat{\rho}]+\frac{\kappa}{2}\mathcal{L}_{a}(\hat{\rho})+\frac{\gamma_{x}}{2}\mathcal{L}_{\hat{\sigma}}(\hat{\rho})+\frac{P_{x}}{2}\mathcal{L}_{\hat{\sigma}^{\dagger}}(\hat{\rho})\notag \\ &+&\frac{P_{\theta}}{2}\mathcal{L}_{\hat{\sigma}\hat{a}^{\dagger}}(\hat{\rho}).
\end{eqnarray}
The Lindblad superoperator $\mathcal{L}_{\hat{X}}$ is defined for an arbitrary operator $\hat{X}$ as $\mathcal{L}_{\hat{X}}(\hat{\rho})=2\hat{X}\hat{\rho}\hat{X}^{\dagger}-\hat{X}^{\dagger}\hat{X}\hat{\rho}-\hat{\rho}\hat{X}^{\dagger}\hat{X}$. It is worth mentioning that the last term that appears in this master equation together with an additional incoherent decay term $(\gamma_\theta/2)\mathcal{L}_{\hat{\sigma}^{\dagger}\hat{a}}$ have been proposed by Majumdar {\it et al.}~\cite{Majumdar:2010}, as a model for describing phonon-mediated off-resonant QD-cavity coupling. In fact, the term associated with a decay rate $P_\theta$ is being included above into the master equation as a process that corresponds to the creation of a cavity photon together with the collapse of the QD to its ground state. Additionally, this decay process is the most important for observing cavity emission under resonant excitation of the QD as first pointed out in the seminal paper by Majumdar. It is interesting to note that in a previous experimental work on a single QD coupled to a PhCC under resonant excitation, a simple phonon dephasing  model was investigated and suggested that this theoretical approach could be useful for describing the off-resonant QD-cavity interaction~\cite{Englund:2009,Ulhaq:2010}. The other incoherent process given by the Lindblad term $(\gamma_\theta/2)\mathcal{L}_{\hat{\sigma}^{\dagger}\hat{a}}$ that describes the annihilation of a cavity photon with one excitation of the QD can be safely neglected for our purposes, since this process is strongly attenuated when the QD on average remains excited as is the case when a system is operating under the conditions $P_x>\gamma_x\sim0$.  We have confirmed that the presence of this term with a small decay rate $\gamma_{\theta}$ do not affect qualitatively our findings.  It is noteworthy that the decay rate $P_\theta$ corresponds to a mechanism that depends on the temperature, but differs qualitatively from the pure dephasing model owing to the fact that the presence of cavity photons affects the QD linewidth. In fact, this mechanism produces an inherent asymmetry between absorption and emission rates of phonons in the system, and it can be well described by assuming that $P_\theta$ behaves like an S-shaped curve. It is,
\begin{equation}\label{ptheta}
P_{\theta}=\frac{\tilde{P}_{\theta}}{1+Ae^{-B(T-T')}}
\end{equation}
with $\tilde{P}_{\theta}$, $A$, $B$ and $T'$ parameters that incorporates the whole effect of many-body effects as the electron-phonon interaction, the spectral density and the mean number phonons of the environment, as well as any other possible process related to the influence of the temperature. This particular dependence on the temperature for the decay rate $P_\theta$ is a conceivable hypothesis, since a recent experimental work in electrically tunable single QD nanocavities revealed that the temperature dependence of the pure dephasing rate changes similarly to an S-shaped curve. It is, the pure dephasing rate does not vary for low temperatures, but increases rapidly for higher temperatures. Additionally, it was concluded that the temperature dependence on the dephasing rate is an evidence for decoherence mediated by coupling to acoustic phonons~\cite{Laucht:2009}. The temperature dependence for $P_\theta$ given by the Eq.~(\ref{ptheta}) is a phenomenological proposal that could be tested experimentally and do not affects the main results of this paper. Moreover, we have considered the temperature explicitly instead of the detuning, since the thermal dependence of the frequencies in PhCC with embedded QDs is incorporated through the refractive index~\cite{Blakemore:1982} and the Varshni model~\cite{Varshni:1967,Vurgaftman:2001}. More precisely,  $\omega_c(T)=\omega_c(0)/(1+aT)$ and   $\omega_x(T)=E_g(0)-\alpha T^2/(T+\beta)$. In particular, we have chosen the parameters $\omega_c(0)=1043.27meV$, $a =0.852\times10^{-5}K^{-1}$,  $E_g(0)=1044.5\,meV$, $\alpha=0.7\,meV K^{-1}$ and $\beta=590\,K$ for comparison purposes with the experimental setup used by Tawara {\it et al.} in Ref.~\cite{Tawara:2010}. In what follows, we are interested in the emission properties of the system, and therefore we compute the photoluminescence spectrum (PL)  based on the quantum regression formula (QRF)~\cite{Perea:2004}. Formally, it corresponds to compute $S(\omega)=\int_{-\infty}^{\infty}\meanv{\hat{a}^{\dagger}(\tau)\hat{a}(0)}e^{-i\omega \tau}d\tau$ where the quantity $\meanv{\hat{a}^{\dagger}(\tau)\hat{a}(0)}$ is the two-time correlation function of the cavity field. In Fig.~\ref{Fig-Tawara}(a) the emission spectra under low pumping regime $P_x/(\kappa+\gamma_x)<1$ is shown as a function of the temperature where the two emission peaks are labeled as ($C$) and ($X$), respectively. 
This identification is done only for comparison purposes with the reference mentioned above. However, this issue is more fundamental than has been pointed by the authors and it deserves special attention since there are fundamental physics behind it.  We will go back to this issue when discussing the effect of phonon-mediated coupling over the optical transitions of the system. The Fig.~\ref{Fig-Tawara}(b) shows the spectral peak positions as a function of the temperature, where the  peak $C$ presents a blueshift confirming the attraction phenomenon from cavity mode toward the exciton mode.
\subsection{Description based on Lindblad master equation approach without gain.}\label{lindbladWG}
\begin{figure}[ht!]
\centering
\includegraphics[scale=1.0]{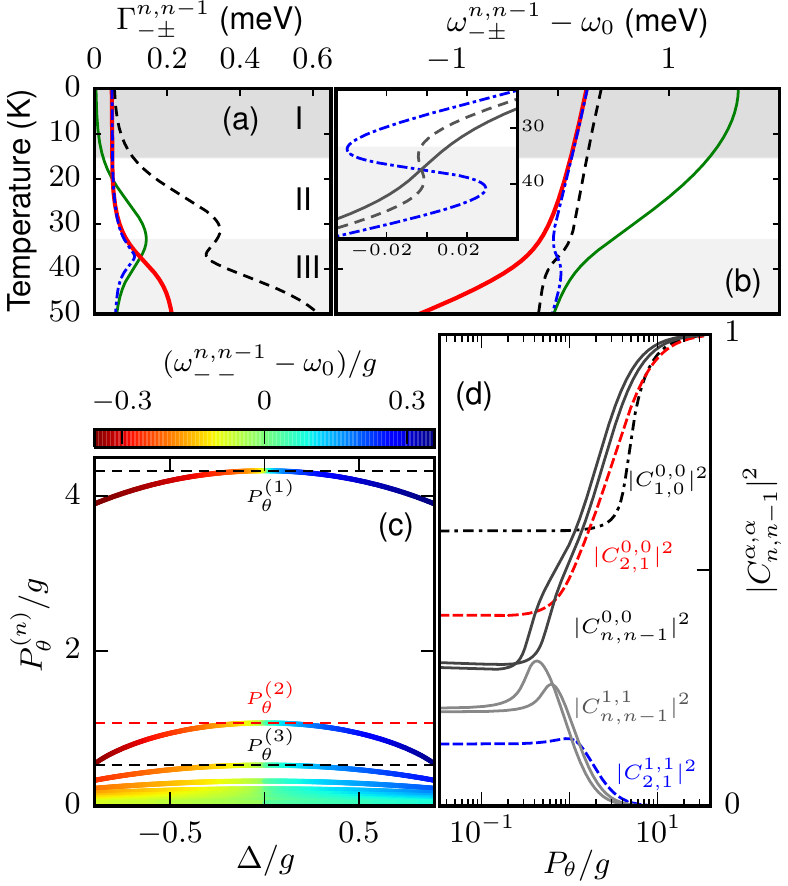}
\caption{Panel (a) shows the linewidth $\Gamma^{1,0}_{-\,-}$ (solid green line), $\Gamma^{1,0}_{-\,+}$ (thick red  line), $\Gamma^{2,1}_{-\,-}$ (dot-dashed blue line) and $\Gamma^{2,1}_{-\,+}$ (dashed black line) as a function of the temperature. Moreover, three distinct regions (see text above) are shown in the temperature range. Panel (b) shows the spectral peak positions $\omega^{n,n-1}_{-\,\pm}$ for $n=1,2$ and the same color convention is adopted within the panel (a). The inset shows $\omega^{n,n-1}_{-\,-}$ for $n=2,3,4$ with blue dot-dashed, gray dashed and solid gray line, respectively. Panel (c) shows the distribution of $P^{(n)}_{\theta}$ whereas the detuning has been changed closely at resonance frequency. Furthermore, the color-code specifies the spectral position that each singlet has when it begins to contribute to the resonance state. Panel (d) shows the coefficient $\left|C^{0,0}_{n,n-1}\right|^2$ for $n=1$ (dot-dashed black line), $n=2$ (red dashed line) and  for higher rungs $n=3,4$ (dark gray line). Moreover, the coefficient $\left|C^{1,1}_{n,n-1}\right|^2$ for $n=2$ (blue dashed line) and for higher rungs $n=3,4$ (light gray line). The parameters chosen in our numerical simulations are: $\gamma_x/g=0.003$, $\kappa/g=0.33$, (a)-(b) $g=0.3\,meV$, $P_\theta/g=1.08$ $A=0.5$, $B=0.2\,K^{-1}$, $T'=30\,K$, $\tilde{P}_{\theta}/g=1.5$ (d) $\Delta/g=0.33$. (For interpretation of the references to color in this figure legend, the reader is referred to the web version of this article.)
}\label{Fig-2}
\end{figure}

\begin{figure}[h!]
\begin{center}
\includegraphics[scale=1.]{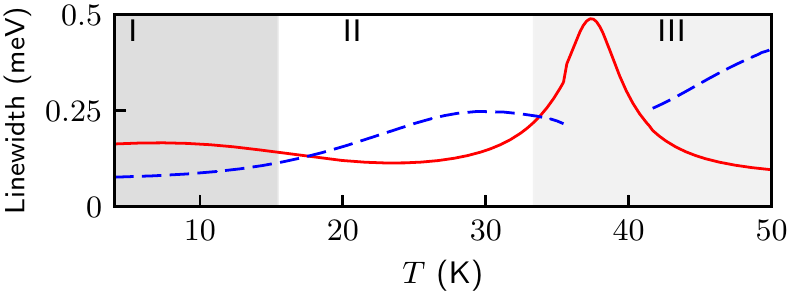}
\caption{Numerical calculation of the linewidths of the two peaks shown in Fig.~\ref{Fig-Tawara}(a) as a function of temperature. The peak linewidth of $C$ and $X$ are shown as solid red and dashed blue line, respectively. (For interpretation of the references to color in this figure legend, the reader is referred to the web version of this article.)}\label{fig-3}
\end{center}
\end{figure}

While the master equation approach given by the Eq.~(\ref{eq1}) shows a good qualitative agreement with the experimental results reported in Fig.~$2$ (a)-(b) from Ref.~\cite{Tawara:2010}, it is difficult to achieve manageable dynamical equations for  
extracting quantitative information on the fundamental physics related with the phonon-mediated coupling, and the role played by the phonons in the cavity-to-exciton attraction phenomenon. Therefore, we consider a simplified as well as accurate master equation given by $d\hat{\tilde{\rho}}/dt=-i[\hat{K},\hat{\tilde{\rho}}\,]+P_{\theta}\mathcal{L}_{\hat{\sigma}\hat{a}^{\dagger}}(\hat{\tilde{\rho}}\,)/2\equiv{\cal L}\hat{\tilde{\rho}}$ where the losses are incorporated in an effective manner with $\hat{K}=\hat{H}-i\gamma_{x}\hat{\sigma}^{\dagger}\hat{\sigma}/2-i\kappa\hat{a}^{\dagger}\hat{a}/2$ and $P_x=0$. A remarkable aspect of this master equation, is that it accounts for the same fundamental physics as the Eq.~(\ref{eq1}) and the number of excitation $\hat{N}_{exc}=\hat{a}^{\dagger}\hat{a}+\hat{\sigma}^{\dagger}\hat{\sigma}$ commutes with $\hat{H}$. Moreover, the eigenvalues of $\hat{H}$ defines the rungs in the JC ladder. Interestingly enough, $[\mathcal{L}_{\hat{\sigma}\hat{a}^{\dagger}},\hat{N}_{exc}]=0$ implies that the Liouvillian ${\cal L}$ can be partitioned into subspaces that can be written in terms of $4\times4$ matrices. As an immediate consequence of this, the $n$th subspace that describes one-photon transitions reads
\begin{equation}
\mathcal{L}^{n,n-1}=
\begin{bmatrix}
\frac{\kappa-(2n-1)P_\theta}{2} & -i\Omega_{n-1} & i\Omega_{n} & 0 \\
-i\Omega_{n-1} & \frac{\gamma_x+Z}{2} & 0 & i\Omega_{n}   \\
i\Omega_{n} & 0 & \frac{2\kappa+P_\theta+Z^*-\gamma_x}{2} &-i\Omega_{n-1}   \\
    \sqrt{n(n-1)}P_\theta & i\Omega_{n} & -i\Omega_{n-1} & \frac{\kappa}{2}
\end{bmatrix}
\end{equation}
and it becomes accompanied with the eigenvalue problem $\mathcal{L}^{n,n-1}{\mathbf U}^{n,n-1}=\lambda^{n,n-1}{\mathbf U}^{n,n-1}$, where $\lambda^{n,n-1}$ and ${\mathbf U}^{n,n-1}$ denotes their eigenvalues and eigenvectors, respectively. We used a shorthand notation for $\Omega_n=g\sqrt{n}$ and $Z=-nP_{\theta}+4i\Delta$ (and $Z^*$ denoting its complex conjugate). Also $\lambda^{n,n-1}$ has been used for identifying four distinct eigenvalues, namely  $\lambda^{n,n-1}_{\p\p},\, \lambda^{n,n-1}_{\p\m},\,\lambda^{n,n-1}_{\m\p}$ and $\lambda^{n,n-1}_{\m\,\m}$. This theoretical approach has been introduced in the past for finding solutions of Lindblad master equations without gain~\cite{Torres:2014}. A detailed analysis on the spectral peak positions $\omega^{n,n-1}_{\pms\,\pms}=\text{Im}[\lambda^{n,n-1}_{\pms\,\pms}]$ and the linewidths $\Gamma^{n,n-1}_{\pms\,\pms}=\text{Re}[\lambda^{n,n-1}_{\pms\,\pms}]$ of the $n$th subspace, states that two of the four optical transitions reinforces the background emission as a consequence of the effect of $P_{\theta}$, whereas the eigenvalues $\lambda^{n,n-1}_{\m\,\pms}$ are related with transitions that significantly contributes to the emission spectrum of the system. For a temperature range of $0-15K$ (region I), it can be assumed that $P_{\theta}$ takes a negligible value ($P_{\theta}/\tilde{P}_{\theta}\lesssim 0.1$) and the off-resonance coupling will not plays an important role in the dynamics. Hence, the emission properties of the system will corresponds to the well-known JC model. While $P_{\theta}$ increases ($0.1\lesssim P_{\theta}/\tilde{P}_{\theta}\lesssim0.8$) in a temperature range of $15-33K$ (region II), it will cause a selective broadening  in the linewidths $\Gamma^{n,n-1}_{\m\,\p}$ on each subspace, it being stronger for higher rungs of the JC ladder. Interestingly, the effect induced by $P_{\theta}$ is completely different when it pass a certain critical value $P_{\theta}/\tilde{P}_{\theta}\gtrsim 0.8$ (region III, for temperatures greater than $33K$), and all the linewidths $\Gamma^{n,n-1}_{\m\,\m}$ will have almost the same value. This phenomenology on the linewidths $\Gamma^{n,n-1}_{\m\,\pms}$ as a function of the temperature is shown in Fig.~\ref{Fig-2}(a). It is illustrated for the particular case of $\Gamma^{1,0}_{\m\,\m}$ ($\Gamma^{1,0}_{\m\,\p}$) with solid green (thick red) line, and also for  $\Gamma^{2,1}_{\m\,\m}$ ($\Gamma^{2,1}_{\m\,\p}$) with dot-dashed blue (dashed black) line, respectively. Notice that the region I, II and III mentioned above are depicted as dark-gray, white and light-gray stripes, respectively. In Fig.~\ref{Fig-2}(b) the spectral peak positions $\omega^{n,n-1}_{\m\,\pms}$ corresponding to $n=1,2$ are shown and the same color convention as panel~(a) has been used. Interestingly enough is the region III, where signatures of strong and weak coupling regimes are evidenced simultaneously in the system. In fact, an anticrossing between the lower and upper polariton of the first rung is observed. Besides, at the resonant frequency an emission peak at $\omega_0$ appears as known in the weak coupling regime. A remarkable aspect of this emission peak is that it exhibits a blueshift toward the QD resonance while the temperature changes, and this spectral shifting is governed by all spectral peak positions $\omega^{n,n-1}_{\m\,\m}$, except for $n=1$. The inset of Fig.~\ref{Fig-2}(b) shows few particular cases $n=2,3,4$ with blue dot-dashed, gray dashed and solid gray line, respectively. This emission peak corresponds to a resonance state \cite{Jung:1999} of small width which is induced by the phonon reservoir. More precisely, this collective state is created as the result of the successive overlapping of the singlet that begins to arise once the eigenvalues $\lambda^{n,n-1}_{\m\,\pms}$ approach each other at a critical value $P^{(n)}_{\theta}$ or so-called exceptional point (EP). This phenomenology is typical of a dynamical phase transition (DPT) in the system~\cite{Rotter:2010}, whose characteristic is the coexistence of both the strong and weak coupling regimes in cQED. It was recently reported as an explanation to the unexpected off-resonance cavity mode emission~\cite{Echeverri-Arteaga:2017}. The distribution of the EPs whereas the detuning has been changed closely at resonance frequency is shown in Fig.~\ref{Fig-2}(c). Each curved-line depicts the set of $P^{(n)}_{\theta}$ for a particular subspace and the color-code specifies the spectral position that  each singlet has when it begins to contribute to the resonance state. Each $P^{(n)}_{\theta}$ defines an straight line (dashed horizontal line) that is tangent to only one curved-line when $\Delta=0$, it implies that the resonance state is formed by the contribution of a singlet from each subspace at frequency $\omega_0$, thus a resonance trapped state is reached~\cite{Heiss:1998}. Under off-resonant condition  $\Delta\neq0$ all the $P^{(n)}_{\theta}$ shift to lower values and it implies that the resonance state is now formed by a singlet from each subspace at frequencies close to $\omega_0$, {\it i.e.} this resonance state is being red-shifted or blue-shifted respect to the crossover energy.
To understand the emission properties of the system during the DPT, we turn our attention to the expansion coefficients $C^{\alpha,\beta}_{n,n-1}$ of the eigenvector ${\mathbf U}^{n,n-1}_{\m\,\m}=\sum_{\alpha,\beta}C^{\alpha,\beta}_{n,n-1}\ket{n-\alpha,\alpha}\bra{n-1-\beta,\beta}$ in the bare-states basis $\big\{\ket{n-\alpha,\alpha}\equiv\ket{n}\vert_{n=0}^{\infty}\otimes\ket{\alpha}\vert_{\alpha=0}^{1}\big\}$. Here $n$ represents the number of photons in the cavity and $\alpha$ the state of the QD. Particularly, the coefficients $C^{0,0}_{n,n-1}$ and $C^{1,1}_{n,n-1}$ are related with the leakage of cavity photons when the QD is in its ground state or its excited state, respectively. The square modulus $\abs{C^{0,0}_{n,n-1}}^2$ and $\abs{C^{1,1}_{n,n-1}}^2$ are shown in Fig.~\ref{Fig-2}(d) for $n=1,2,3,4$. It can be seen that for $P_\theta<P^{(2)}_\theta$ all the optical transitions of the system corresponds to the linear JC model, since these coefficients are practically constant as the case $P_\theta=0$. When $P_\theta>P^{(1)}_\theta$ all the rungs (each subspace) of the JC ladder contributes with a singlet, such that its coefficients collectively behaves as $C^{1,1}_{n,n-1}\sim0$ and $C^{0,0}_{n,n-1}\sim1$. It indicates that the phonon-mediated coupling is decoupling the QD-cavity system with the QD conditioned to be in its ground state, together with the complete formation of the resonance state. This state has emission properties similar to a QD-Cavity system in the weak coupling regime with cavity pumping and without excitonic pumping. When the phonon-mediated coupling is in the range $P^{(2)}_\theta<P_\theta<P^{(1)}_\theta$, the emission properties of the system will be characterized by the presence of the Rabi splitting of the first rung (lower than $2g$), as a remainder of the strong coupling regime, and a singlet that clearly blueshift toward the exciton resonance. This blueshift is an immediate consequence of the DPT in the system, where characteristics of strong and weak coupling regimes are shared simultaneously. This phenomenology can be coined as the {\it quantum intermediate coupling regime} in agreement with the hypothesis of Tawara's work. Finally, we discuss the relevance of our findings for understanding why the identification of the peaks with labels $C$, $X$ used by Tawara et al. and also introduced by us for comparison purposes in Fig.~\ref{Fig-Tawara} is wrong. The emission peak $C$ and $X$ cannot be associated with the bared cavity and the QD, since the system do not operates in a genuine weak coupling regime. Instead of this, the system is in an intermediate quantum regime where the apparent crossing observed is the result of the formation of a resonance state. Hence, the emission peak $C$ corresponds to a collective emission from all rungs in the JC ladder near to the cavity frequency  and the emission peak $X$ is due to optical transitions from the two polaritons of the first rung. While theoretically the pronounced spectral shifting in the experimental measurements is well-accounted, additional information from the experimental side is required to confirm the phenomenology described here. We have good reasons to believe that new experimental results addressed on the linewidths could display similar changes in the linewidths as shown in Fig.~\ref{fig-3}. It will be expected an inversion on the linewidths (broadening-narrowing) far off-resonance, together with a significant broadening on the linewidth of the peak labeled $C$ when the system is close to the resonance. 
%
\section{Conclusions}\label{sec:conclusions}
The quantum phenomenon reported experimentally by Tawara {\it et al}. and known as the strange cavity-to-exciton attraction has been successfully explained in this investigation. Within the framework of a Lindblad master equation approach, we have obtained accurate and comprehensive results demonstrating that the phonon-mediated coupling originates the cavity mode attraction toward the exciton mode, moreover this phenomenology is related with a dynamical phase transition in the system. In fact, in this quantum domain the system does not strictly operate in weak or strong coupling regime, but in an intermediate coupling regime which is characterized by the emission properties of the linear JC model together with the formation of a resonance state close to the cavity frequency. 
\section*{Acknowledgements}
\noindent
S.E.-A. and H.V.-P. gratefully acknowledge funding by
COLCIENCIAS under the project  ``Emisi\'on en sistemas de Qubits
Superconductores acoplados a la radiaci\'on''. C\'odigo 110171249692, CT 293-2016, HERMES 31361 and the  project  ``Interacci\'on radiaci\'on-materia mediada por fonones en la electrodinámica cu\'antica de cavidades''. C\'odigo HERMES 42134. S.E.-A. also acknowledges support from the  ``Beca de
Doctorados Nacionales de COLCIENCIAS 727''. E.A.G acknowledges financial support from
Vicerrector\'ia de Investigaciones of the Universidad del Quind\'io through the project No. 919.

\begin{thebibliography}{1}
\bibitem{Reithmaier:2004} J. P. Reithmaier, G. S\c{e}k, A. L\"offler, C. Hofmann, S. Kuhn, S. Reitzenstein, L. V. Keldysh, V. D. Kulakovskii, T. L. Reinecke, A. Forchel, Strong coupling in a single quantum dot-semiconductor microcavity system, Nature. 432 (2004) 197-200.
%
\bibitem{Shudong:1998} J. Shudong, M. Susumu, Y. S. Takiguchi, Y. Yamamoto, Direct time-domain observation of transition from strong to weak coupling in a semiconductor microcavity, Appl. Phys. Lett. 73 (1998) 3031--3033 (1998).

\bibitem{Bulovic:1998} V. Bulovi\'c, V. B. Khalfin, G. Gu, P. E. Burrows, D. Z. Garbuzov, S. R. Forrest, Weak microcavity effects in organic light-emitting devices, Phys. Rev. B. 58 (1998) 3730--3740.

\bibitem{Fletcher:2000} R. B. Fletcher, D. G. Lidzey, D. D. C. Bradley, Spectral properties of resonant-cavity, polyfluorene light-emitting diodes, Appl. Phys. Lett. 77 (2000) 1262--1264.

%
\bibitem{Salter:2010} 
C. L. Salter, R. M. Stevenson, I. Farrer, C. A. Nicoll, D. A. Ritchie, A. J. Shields, An entangled-light-emitting diode, Nature 465 (2010) 594--597.
\bibitem{Moreau:2001} E. Moreau, I. Robert, J. M. Gérard, I. Abram, L. Manin, V. Thierry-Mieg, Single-mode solid-state single photon source based on isolated quantum dots in pillar microcavities, Appl. Phys. Lett. 79 (2011) 2865--2867.

\bibitem{Tsintzos:2008} S. I. Tsintzos, N. T. Pelekanos, G. Konstantinidis, Z. Hatzopoulos, P. G. Savvidis, A GaAs polariton light-emitting diode operating near room temperature, Nature. 453 (2008) 372--375.

\bibitem{Wei:2014} H.-R. Wei, F.-G. Deng, Scalable quantum computing based on stationary spin qubits in coupled quantum dots inside double-sided optical microcavities, Sci. Rep. 4 (2014) 7551--7559.

\bibitem{Ginzburg:2016} P. Ginzburg, Cavity quantum electrodynamics in application to plasmonics and metamaterials, Rev. Phys. 1 (2016) 120--139.

\bibitem{Kiraz:2003} 
 A. Kiraz, C. Reese, B. Gayral, L. Zhang, W. V. Schoenfeld, B. D. Gerardot, P. M. Petroff, E. L. Hu, A. Imamoglu, Cavity-quantum electrodynamics with quantum dots, J Opt B Quantum Semiclassical Opt. 5 (2013) 129--137.
%
\bibitem{Tawara:2010} T. Tawara, H. Kamada, T. Tanabe, T. Sogawa, Cavity-QED assisted attraction between a cavity mode and an exciton mode in a planar photonic-crystal cavity, Opt. Express 18 (2010) 2719-2728.
%
\bibitem{Valente:2014} D. Valente, J. Suffczy\ifmmode \acute{n}\else \'{n}\fi{}ski, T. Jakubczyk, A. Dousse, A. Lema\^{\i}tre, I. Sagnes, L. Lanco, P. Voisin, A. Auff\`eves,  P. Senellart, Frequency cavity pulling induced by a single semiconductor quantum dot, Phys. Rev. B 89 (2014) 041302--041307.

\bibitem{Hennessy:2007} K. Hennessy, A. Badolato, M. Winger, D. Gerace, M. Atat\"ure, S. Gulde, S. F\"alt, E.L. Hu, A. Imamo\u glu, Quantum nature of a strongly coupled single quantum dot–cavity system, Nature 445 (2007) 896--899.
%
\bibitem{Kaniber:2008} M. Kaniber, A. Laucht, A. Neumann, J. M. Villas-B\^{o}as, M. Bichler, M.-C. Amann, J. J. Finley, Investigation of the nonresonant dot-cavity coupling in two-dimensional photonic crystal nanocavities, Phys. Rev. B 77 (2008) 161303--161307(R).
%
\bibitem{Echeverri-Arteaga:2017} S. Echeverri-Arteaga, H. Vinck-Posada, and E. A. G\'omez, Explanation of the quantum phenomenon of off-resonant cavity-mode emission, Phys. Rev. A 97 (2018) 043815--043821.
%
\bibitem{Majumdar:2010} A. Majumdar, E.D. Kim, Y. Gong, M. Bajcsy, J. Vu\u ckovi\'c, Phonon mediated off-resonant quantum dot–cavity coupling under resonant excitation of the quantum dot, Phys. Rev. B 84 (2011) 085309--085316.
\bibitem{Englund:2009} D. Englund, A. Majumdar, A. Faraon, M. Toishi, N. Stoltz, P. Petroff, J Vu\u{c}kovi\'{c}, Resonant excitation of a quantum dot strongly coupled to a photonic crystal nanocavity, Phys. Rev. Lett. 104 (2010) 073904--073908.

\bibitem{Ulhaq:2010} A. Ulhaq, S. Ates, S. Weiler, S. M. Ulrich, S. Reitzenstein, A. L\"offler, S. H\"ofling, L. Worschech, A. Forchel, P. Michler, Linewidth broadening and emission saturation of a resonantly excited quantum dot monitored
via an off-resonant cavity mode, Phys. Rev. B 82 (2010) 045307--045312.

\bibitem{Laucht:2009} A. Laucht, N. Hauke, J. M. Villas-B\^oas, F. Hofbauer, G. B\"ohm, M. Kaniber, J. J. Finley, Dephasing of exciton polaritons in photoexcited InGaAs quantum dots in GaAs nanocavities, Phys. Rev. Lett. 103 (2009) 087405--087409.

\bibitem{Blakemore:1982} J.S. Blakemore, Semiconducting and other major properties of gallium arsenide, J. Appl. Phys.  53 (1982) R123--R181.
\bibitem{Varshni:1967} Y.P. Varshni, Temperature dependence of the energy gap in semiconductors, Physica 34 (1967) 149--154.
\bibitem{Vurgaftman:2001} I. Vurgaftman, J.R. Meyer, L.R. Ram-Mohan, Band parameters for III-V compound semiconductors and their alloys, J. Appl. Phys. 89 (2001) 5815--5875.

\bibitem{Perea:2004} J. I. Perea, D. Porras, C. Tejedor, Dynamics of the excitations of a quantum dot in a microcavity, Phys. Rev. B. 70 (2004) 115304--115317.

\bibitem{Torres:2014} J.M. Torres, Closed-form solution of Lindblad master equations without gain, Phys. Rev. A. 89 (2014) 052133--052141.

\bibitem{Jung:1999} C. Jung, M. M\"uller, I. Rotter, Phase transitions in open quantum systems, Phys Rev. E. 60 (1999) 114--131.

\bibitem{Rotter:2010} I. Rotter, Dynamical Phase Transitions in Quantum Systems, J. Mod. Phys. 1 (2010) 303--311.

\bibitem{Heiss:1998} W. D. Heiss, M. M\"uller, I. Rotter, Collectivity, phase transitions, and exceptional points in open quantum systems, Phys. Rev. E. 58 (1998) 2894--2901.

\end{thebibliography}

\end{document}